\documentstyle[prl,twocolumn,aps]{revtex}
\input psfig
\begin{document}
\title{Excess Resistance Effect in a Normal Metal Contacting
a Superconductor}
\author{A. Kadigrobov\cite{AK}}
\address{B Verkin Institute for Low  Temperature Physics \&
Engineering,\\ National Academy of Science of Ukraine, 47 Lenin
Av., 310164,  Kharkov, Ukraine}
\author{R. Shekhter and M. Jonson}
\address {Department of Applied Physics, 
Chalmers University of Technology and G{\"o}teborg
University, S-412 96 G{\"o}teborg, Sweden }
\maketitle
\begin{abstract}
In relatively pure normal samples contacting a superconductor we
consider the excess resistance effect (that is a decrease of the
total electrical resistance of the sample after transition of the
superconducting part into the normal state) and determine
conditions under which the effect arises.
\end{abstract}

\vspace{10mm}




 \section*{1. Introduction}
     Recently there have been many  theoretical  and  experimental
investigations of transport properties of systems with mixed normal
and superconducting elements where new effects have been discovered.
Peculiar interplay of the phase coherence intrinsic to the
superconductor and the one in the normal metal on a mesoscopic length
scale gives rise to new effects both in mesoscopic
(see, e.g., \cite{Ptr,Hekk,Volk,Naz,Been} and macroscopic samples.
One of the effects in macroscopic samples is a decrease in the total
electrical resistance
after transition of the superconducting part of the sample into the
normal state  under  the  critical  electric  current  or  critical
magnetic field \cite{AMK,Tz1,Obol,Kad,Kast}. As  a  result,  in
the first  case the current-voltage characteristic of  the  sample
becomes S-shaped  and near  the  superconducting  critical  current
self-oscillations   of  the  current  and  electric   field   arise
\cite{AMK,Tz2}.

Here we consider situations of relatively pure composite samples
with both weak and strong normal scattering of electrons at the
boundaries of two conductors. Such materials have recently been
obtained  \cite{Nit}.

It occurs that  there  are  several  physical mechanisms that work
differently in different physical situations, but they all result
in the behavior of the resistance mentioned above.
A physical description of the mechanisms and determination of the
excess resistance effect in macroscopic samples is performed in
Section 2. A solution of
Boltzmann's equation and determination of the resistance in general
terms of the probabilities of the Andreev and normal reflections at
the N-S boundary is given in Appendix.
\section*{2. Excess resistance effect in kinetics}
      a)  Contact of a semiconductor and a superconductor  with   a
negligibly low Schottky barrier.
It is known that electrons incident from a normal conductor  to  an
N-S boundary undergo Andreev reflection at it and the boundary does
not contribute to the resistance of the sample (here and  below  we
neglect the excess resistance emerging as a result  of  penetration
of electric field into the superconductor). However, the  electrons
incident on the N-S boundary at small angles do not undergo Andreev
- type, but specular reflection as at an  ordinary  boundary  of  a
conductor \cite{Dzh}. As these electrons do not penetrate the   N-S
boundary, an excess resistance of the N-S boundary  appears.  Their
relative number is $\sim\sqrt{kT / \epsilon_F}$  \cite{AMK}  (k  is
the Boltzman constant, T is temperature, $ \epsilon_F$ is the Fermi
energy). For regular metals this parameter is too small to have  an
impact on the transport properties. If, however,  a  superconductor
is in contact with a semiconductor (or a semimetal), this parameter
increases by several orders of  magnitude  since  it  contains  the
Fermi  energy  of  the  normal  conductor   and,   therefore,   can
qualitatively  modify  kinetic  properties  of  the   semiconductor
contacting a superconductor.

\begin{figure}
\vspace*{5mm}
\centerline{\psfig{figure=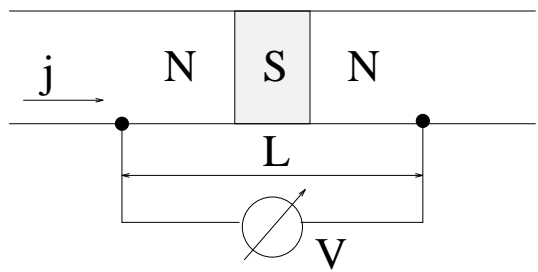,width=7cm}}
\vspace*{5mm}
\caption{ \protect{\label{fig:geometry1}}
Normal conductor(N)-Superconductor(S)-Normal conductor(N)
sample. $I$ is the total transport current through the  sample, $L$  is
the distance between the points where the voltage $V$ is measured.
}
\end{figure}

Here we consider the resistance of a semiconductor - superconductor
- semiconductor system schematically shown in Fig.1. We consider the
case of Schottky barrier absence and  assume  the  mean  free  path
to be $l_0 \ll L$.

In the momentum space $\bf p$ inside a belt of
the width $\delta{ \bf p} = \sqrt{2m\Delta(T)}$ ($m$ is the mass of the
electron, $\Delta(T)$ is the superconducting energy gap) and the
thickness $\Delta(T)/v_F$ ($v_F$ is the Fermi velocity) embracing the
Fermi-sphere and parallel to the N-S boundary, the probability of the
specular reflection at the N-S boundary $T_{\bf p} = 1$ and  $T_{\bf
p} = 0$   outside it \cite{Dzh,sound}. Using this fact and Eq. (A6)
one gets
\begin{equation}
\alpha \equiv \frac{\delta R}{R} =
\left\{ 
\begin{array}{ll} 
\left(\frac{kT}{\epsilon_{F}}\right)^{2}\frac{l_{0}}{L}
& \mbox{$kT \ll \Delta (T)$}\\
1.45\frac{l_{0}}{L} \frac{\Delta_{0}}{kT_{c}}
\left(\frac{\Delta_{0}}{\epsilon_{F}}\right)^{2}
\left(\frac{T_{c}-T} {T_{c}}\right)^{\frac{3}{2}} & 
\mbox{$\Delta (T) \ll kT_{c}$}\\
\sim \left( \frac{l_{0}}{L}\right) 
\left(\frac{\Delta_{0}}{\epsilon_{F}}\right)^{2}
&\mbox{$\Delta (T) \sim kT_{c}$} 
\end{array}\right.
\label{semi} \end{equation}
where
$\frac{\delta R}{R}=(V(L)/I-R)/R $ is the relative  resistance  of
the  N-S boundaries, $R$ is the total resistance of the sample  in
the normal state, $I$ is the total current through the sample,
$\Delta (T)=1.8~\Delta_{0}~(\frac{T_{c}-T}{T_c})^{1/2}$,~
$\Delta_0$ is the superconducting gapwidth at  $T=0$, and $T_c$ is
the critical temperature. As $\alpha >0$, the effects  mentioned
above arise in the situation considered.

\begin{figure}
\vspace*{5mm}
\centerline{\psfig{figure=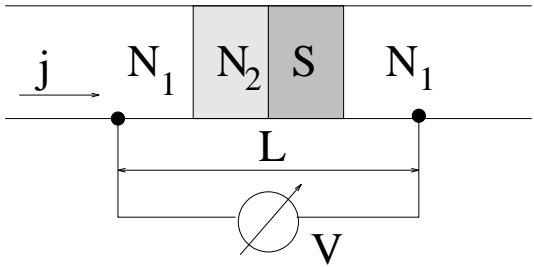,width=7cm}}
\vspace*{5mm}
\caption{ \protect{\label{fig:geometry2}}
A sample with a boundary between two normal  metals  $N_1$
and $N_2$, and two N-S boundaries. $I$  is  the total transport  current
through  the  sample, $L$  is the distance between the points where
the voltage $V$ is measured 
}
\end{figure}

   b) Regular normal metal with a twin or grain boundary contacting
a superconductor.
Here we assume the distance  between the N-N and N-S boundaries  to
be $l < \lambda_T  <l_0$ where the normal metal  'coherent  length'
$\lambda_T = \hbar v_F/kT $. An electron incident from metal $N_1 $
(see Fig.2) to the N-N boundary in the absence of superconductivity
undergoes normal  reflection  with  probability  $\rho^2$.  In  the
presence of superconductivity such an electron undergoes repeatedly
both normal reflections at the N-N boundary and Andreev reflections
at the N-S boundary. As a result, multiple coherent reflections  of
the electron arise, and the total probability for the  electron  to
be  reflected by these two boundaries is

\begin{equation} T_{\bf p} =2\rho^2 \frac{1+\cos\phi}
{1+2\rho^2\cos\phi + \rho^4} \label{prob1}
\end{equation}
where $\phi = 2m\epsilon l/\hbar (p_F^2 -p^2)  \sim  l/\lambda_T  $
with $p_F$ the Fermi momentum, $p$ the projection of  the  incident
electron momentum perpendicular to the boundaries,  and  $\epsilon$
is the incident electron energy measured from the Fermi energy. The
probability $T_{\bf p}$ averaged over the incident angles $\phi$ is

\begin{equation}
 <T_{\bf p}> = 2\rho^2/(1+\rho^2)  \label{prob2}
 \end{equation}
If the distance between the N-N and  N-S  boundaries  is  equal  to
zero, the probability $T_{\bf p}$ coincides with the  one  obtained  in
\cite{Klap} when written in terms of $\rho^2$.

Using Eq. (A6), (\ref {prob1}) and (\ref {prob2})  we  have  the
relative  excess resistance $\alpha$ as
\begin{equation}
\alpha = \frac{l_0}{L}(W- \frac{R_S}{R_L}) \label{res2}
\end{equation}
Here $W =<T_{\bf p}>-\rho^2$, and $R_s$ and $R_L$ are  the  resistances
of the S-part of the sample (when it is in the normal state  )  and
the  total  resistance  of  the  N-metal   of   the   length   $L$,
respectively. In  deriving  (\ref{res2})  we  assumed  $\Delta  (T)
\ll\Delta_0$

c)Regular normal metal with impurities contacting a superconductor.
($L \gg l_0 \gg \lambda_T$).
As shown in \cite{H} combined scattering of an electron by an
impurity and an N-S boundary is of a multiple coherent character.
According to \cite{Kad} the cross-section of the electron
back-scattering by an impurity located inside a normal metal layer of thickness
$\sim \lambda_T$ adjoining the N-S is $\bar \sigma = 2\sigma_0$ if
averaged over the distance between the impurity and the N-S boundary
($\sigma_0$ is the cross-section of the scattering by the impurity in
the absence of the N-S boundary). Therefore, this layer as a whole
scatters the electron backwards with the probability
\begin{equation}T^{eff} =  c_i\lambda_T \sigma_0 \label{eff}
\end{equation}  that can be treated as the
effective probability of the
normal reflection by the N-S boundary (the Andreev reflection
probability is $1 - T^{eff}$). Using (\ref{eff}) and Eq. (A6) one
finds the  excess resistance to be \begin{equation}\delta R = \alpha R =
R_N \frac{\lambda_T}{L} - R_S \label{end} \end{equation}

As is evident from (\ref{semi}), (\ref{res2}) and (\ref{end}) the excess resistance $\delta R$
can be positive in many experimental situations and, therefore, in this case
transition of
the superconductor to the normal state is accompanied by a decrease
of the total resistance of the sample.
\section*{Appendix}
Here we find the voltage drop $\delta \Phi$ between the N-S boundary
($x = 0$) and a plane $x = L_0 \gg l_0 $ (for definiteness sake we assume
the normal part of the sample to occupy the right half-space $x >
0$, the coordinate axis $x$ is perpendicular to the N-S boundary) in
the case that a normal metal electron undergoes both the Andreev and
the specular reflection at the N-S boundary with the probabilities
$T_{\bf p}$ and $R_{\bf p}$, respectively ($T_{\bf p} + R_{\bf p} =
1$). Under conditions of a weak
electric field $\it E$ and $l_0 \gg \lambda_T $ the resistance of a
normal conductor  contacting a superconductor is determined by usual
Boltzmann's equation $$ v_{\bf p} \frac{\partial
f_1}{\partial x} + \frac{f_1 - <f_1>_{\epsilon}}{t_0} = e{\it E(x)} v_{\bf p}
\frac{\partial f_0}{\partial E}
\eqno A1 $$
with the boundary condition at the N-S boundary $$f_1^{+}({\bf p}; 0,
y, z) = (T_{\bf p} - R_{\bf p}) f_1^{-}({\bf p}; 0,
y, z)   \eqno A2$$  Here $ v_{\bf p}$ is the x-component of the
electron velocity, $t_0$ is the relaxation
time, $f_1^{+}({\bf p}; {\bf
r})$ and $  f_1^{-}({\bf p}; {\bf r}) $ are nonequilibrium corrections
to the Fermi distribution function $f_0$ for electrons with velocities
directed towards the N-S boundary and away from it, ${\bf r} = (x,y,z)$,
the brackets $ <...>_{\epsilon}$ designate the average over  ${\bf p}$
at a given energy $\epsilon$.

Under the condition of a fixed current $j$ flowing through the system
the electric field $\it E(x)$ in the normal part of the sample is
determined by the local neutrality condition $$dj/dx = 0 \eqno A3$$
Below we find the voltage drop $\delta \Phi$ assuming the normal
reflection probability $T_{\bf p} \ll 1$. Solving Boltzmann's
equation (A1) with the boundary condition (A2) and using
Eq. (A3) one finds the equation for the electric potential $\phi(x)$ in the
normal part of the sample
$$\begin{array}{c}
\int_0^{\infty}\{\frac{e^{-|x -
x'|/l_{\bf p}}}{l_{\bf p}} \}\phi(x')dx' - \int_0^{\infty}\{\frac{e^{-|x +
x'|/l_{\bf p}}}{l_{\bf p}} \}\phi(x') dx'  - 2\phi(x) = \\ 2{\it
E}_{\infty}\{T_{\bf p}  l_{\bf p} e^{-x/l_{\bf p}}\} -
2\phi(0)\{e^{-x/l_{\bf p}}\}
\end{array}  \eqno A4$$
Here $l_{\bf p} = t_0|v_{\bf p}|$, $\{...\} =
<\Theta(v_{\bf p})... >_{\epsilon}$, $\Theta(v_{\bf p}) = 1$
if $v_{\bf p} > 0$ and $\Theta(v_{\bf p}) = 0$
if $v_{\bf p} < 0$,  $\phi(x)$ is associated with $\it E(x)$ by the
relation $${\it E}(x) = R I/L_0 - d\phi/dx \eqno A5$$
where $R$ is the total resistance of the normal conductor of the length
$L_0$ in the absence
of the superconductor, $I$ is the total current, $\phi(x) \to 0$
at $x \to \infty$, $\phi(0)$ is
the value of the electric potential at the N-S boundary ($x = 0$).
The left side of Eq. (A5) is orthogonal to $x$ and the solvability
condition for Eq. (A5) determines  $\phi(0)$ which, together with Eq.
(A5),
gives $$\delta\Phi = I{\it R}(1 + \frac{\{ T_{\bf p} l_{\bf
p}^3\}}{L_0\{l_{\bf p}^2\}}) \eqno A6$$

\section*{Acknowledgment} The work presented in this paper
is supported by INTAS project: N 94-3862.

\end{document}